\documentclass[12pt]{article}
\begin{document}

\begin{center}
{\LARGE On algebraic integrability of the deformed elliptic Calogero--Moser
problem}\vspace{0.7cm}
\end{center}

{\large L. A. Khodarinova}$^{{\dag }}${\large \ and\ I. A. Prikhodsky}$^{{%
\ddag }}${\large \ \vspace{0.3cm}}

$^{{\dag }}$\textit{Magnetic Resonance Centre, School of Physics and
Astronomy, University of Nottingham, Nottingham, England NG7 2RD, e-mail:
LarisaKhodarinova@hotmail.com}

$^{{\ddag }}$\textit{Institute of Mechanical Engineering, Russian Academy of
Sciences,M. Haritonievsky, 4, Centre, Moscow 101830 Russia }

\vspace{1.0cm}

\begin{center}
\textbf{Abstract.}
\end{center}

\begin{quote}
{\small Algebraic integrability of the elliptic Calogero--Moser quantum
problem related to the deformed root systems }$A_{2}(2)${\small \ is proved.
Explicit formulae for integrals are found. }\vspace{0.3cm}
\end{quote}

Following to \cite{Kr} (see also~\cite{ChV} and \cite{VSCh}) we call a Schr%
\"{o}dinger operator 
\[
L=-\Delta +u\left( \vec{x}\right) ,\qquad \vec{x}\in R^{n},
\]%
\textit{integrable} if there exist $n$ commuting differential operators $%
L_{1}=L,$ $L_{2},\ldots ,L_{n}$ with constant algebraically independent
highest symbols $P_{1}\left( \vec{\xi}\right) =\left( \vec{\xi}\right) ^{2},$
$P_{2}\left( \vec{\xi}\right) ,\ldots ,P_{n}\left( \vec{\xi}\right) ,$ and 
\textit{algebraically integrable} if there exists at least one more
differential operator $L_{n+1},$ which commutes with the operators $L_{i},$ $%
i=1,\ldots ,n,$ and whose highest symbol $P_{n+1}\left( \vec{\xi}\right) $
is also independent on $x$ and takes different values on the solutions of
the algebraic system 
\begin{equation}
P_{i}\left( \vec{\xi}\right) =c_{i},i=1,\ldots ,n,  \label{odin}
\end{equation}%
for generic $c_{i}.$

The question how large is the class of the algebraically integrable Schr\"{o}%
\-din\-ger operators is currently far from being understood, so any new
example of such an operator is of substantial interest.

The main result of this paper is the proof of the algebraic integrability of
the following Schr\"{o}dinger operator 
\begin{equation}
L=-\frac{\partial ^{2}}{\partial x_{1}^{2}}-\frac{\partial ^{2}}{\partial
x_{2}^{2}}-m\frac{\partial ^{2}}{\partial x_{3}^{2}}+2(m+1)\left( m\wp
(x_{1}-x_{2})+\wp (x_{1}-x_{3})+\wp (x_{2}-x_{3})\right)  \label{dva}
\end{equation}
when the parameter $m=2.$ Here $\wp $ is the classical Weierstrass elliptic
function satisfying the equation $\wp ^{\prime }{}^{2}-4\wp ^{3}+g_{2}\wp
+g_{3}=0.$ The operator (\ref{dva}) was introduced by Chalykh, Feigin and
Veselov in \cite{VFC} and is related to the deformed root system $\mathbf{%
A_{2}(m)}$ (see \cite{VFC} for details).

When $m=1$ this is the well-known three-particle Calogero--Moser problem.
The usual integrability of this problem has been established by Calogero,
Marchioro and Ragnisco in~\cite{CMR}, the algebraic integrability has been
proved in \cite{Kh} and in a more general case in \cite{BEG}.

The deformed Calogero--Moser\ system~(\ref{dva}) in the trigonometric and
rational limits has been completely investigated by Chalykh, Feigin and
Veselov in \cite{VFC}, where the algebraic integrability for the
corresponding systems has been proved for any $m.$ They have also
conjectured that the same is true in the elliptic case.

In this paper we prove this conjecture for $m=2.$ The first results in this
direction have been found in~\cite{Kh}, where it was proved that problem~(%
\ref{dva}) is integrable. The corresponding integrals have the form

\begin{equation}
\begin{array}{l}
L_{1}=L=-\partial _{1}^{2}-\partial _{2}^{2}-m\partial
_{3}^{2}{}+2(m+1)\left( m\wp _{12}+\wp _{13}+\wp _{23}\right) , \\ 
L_{2}=\partial _{1}+\partial _{2}+\partial _{3}, \\ 
L_{3}=\partial _{1}\partial _{2}\partial _{3}+\left( \frac{1-m}{2}\right)
(\partial _{1}+\partial _{2})\partial _{3}^{2}+\left( \frac{1-m}{2}\right)
\left( \frac{1-2m}{3}\right) \partial _{3}^{3}+ \\ 
\qquad {}+(m+1)\left( \wp _{23}\partial _{1}+\wp _{13}\partial _{2}\right)
+m(m+1)\wp _{12}\partial _{3}+ \\ 
\qquad {}+\left( \frac{1-m}{2}\right) (m+1)\left( (\wp _{13}+\wp
_{23})\partial _{3}+\partial _{3}(\wp _{13}+\wp _{23})\right) ,%
\end{array}
\label{tri}
\end{equation}%
where we have used the notations $\partial _{i}={\partial }/{\partial }x_{i}{%
,}$ $\wp _{ij}=\wp (x_{i}-x_{j}).$

It was also shown that the operator 
\begin{eqnarray}
L_{12} &=&(\partial _{1}-m\partial _{3})^{2}(\partial _{2}-m\partial
_{3})^{2}  \label{chetyre} \\
&&-2(m+1)^{2}\wp _{23}(\partial _{1}-m\partial _{3})^{2}-2(m+1)^{2}\wp
_{13}(\partial _{2}-m\partial _{3})^{2}  \nonumber \\
&&{}+2m(m+1)\left( \wp _{12}-\wp _{13}-\wp _{23}\right) (\partial
_{1}-m\partial _{3})(\partial _{2}-m\partial _{3})  \nonumber \\
&&-m(m+1)\left( \wp _{12}^{\prime }+m\wp _{13}^{\prime }+3(m+1)\wp
_{23}^{\prime }\right) (\partial _{1}-m\partial _{3})  \nonumber \\
&&{}-m(m+1)\left( -\wp _{12}^{\prime }+m\wp _{23}^{\prime }+3(m+1)\wp
_{13}^{\prime }\right) (\partial _{2}-m\partial _{3})  \nonumber \\
&&{}-m(m+1)\wp _{12}^{\prime \prime }-3/2m^{2}(m+1)^{2}\wp _{13}^{\prime
\prime }-3/2m^{2}(m+1)^{2}\wp _{23}^{\prime \prime }  \nonumber \\
&&+m^{2}(m+1)^{2}\left( \wp _{12}^{2}+\wp _{13}^{2}+\wp _{23}^{2}\right)
+2m(m+1)^{2}\left( \wp _{12}\wp _{13}+\wp _{12}\wp _{23}\right)  \nonumber \\
&&{}+2(m+1)^{2}(2m^{2}+3m+2)\wp _{13}\wp _{23}  \nonumber
\end{eqnarray}
commutes with $L_{1},$ $L_{2},$ $L_{3}$ and therefore is an additional
integral of the problem~(\ref{dva}). Unfortunately, this is not enough for
algebraic integrability of the problem~(\ref{dva}) since the highest symbol
of $L_{12}$ is invariant under permutation of $\xi _{1}$ and $\xi _{2}$ and
therefore takes the same values on some of the solutions of the
corresponding system (\ref{odin}).

In this paper we present an explicit formula of one more integral for the
system~(\ref{dva}) related to the deformed root systems $\mathbf{A_{2}(2).}$
This integral together with the previous integrals guarantees the algebraic
integrability of the system (\ref{dva}) in case of $m=2.$

\textbf{Theorem.}\textit{\ The system (\ref{dva}) with }$m=2$ \textit{%
besides the quantum integrals given by~(\ref{tri}) and~(\ref{chetyre}) has
also the following integral }$L_{13}=I+I^{\ast },$\textit{\ where} 
\begin{eqnarray}
I &=& 
\begin{array}{l}
\frac{1}{2}(\partial _{1}-\partial _{2})^{4}(\partial _{1}-2\partial
_{3})^{2}%
\end{array}
\label{5} \\
&& 
\begin{array}{l}
-9\wp _{13}(\partial _{1}-\partial _{2})^{4}-24\wp _{12}(\partial
_{1}-\partial _{2})^{2}(\partial _{1}-2\partial _{3})^{2}%
\end{array}
\nonumber \\
&& 
\begin{array}{l}
-6(\wp _{12}+\wp _{13}-\wp _{23})(\partial _{1}-\partial _{2})^{3}(\partial
_{1}-2\partial _{3})%
\end{array}
\nonumber \\
&& 
\begin{array}{l}
+\left( \rule[-0.09in]{0in}{0.26in}414\wp _{12}\wp _{13}+18\wp _{12}\wp
_{23}+18\wp _{13}\wp _{23}\right.%
\end{array}
\nonumber \\
&&\qquad \qquad \qquad 
\begin{array}{l}
\left. +72\wp _{12}^{2}+108\wp _{13}^{2}+36\wp _{23}^{2}-\frac{201}{2}g_{2}%
\rule[-0.09in]{0in}{0.26in}\right) (\partial _{1}-\partial _{2})^{2}%
\end{array}
\nonumber \\
&& 
\begin{array}{l}
+\left( \rule[-0.09in]{0in}{0.26in}144\wp _{12}\wp _{13}-144\wp _{12}\wp
_{23}\right.%
\end{array}
\nonumber \\
&&\qquad \qquad \qquad 
\begin{array}{l}
\left. +432\wp _{12}^{2}+18\wp _{13}^{2}-18\wp _{23}^{2}+33g_{2}%
\rule[-0.09in]{0in}{0.26in}\right) (\partial _{1}-\partial _{2})(\partial
_{1}-2\partial _{3})%
\end{array}
\nonumber \\
&& 
\begin{array}{l}
+(288\wp _{12}^{2}-69g_{2})(\partial _{1}-2\partial _{3})^{2}-369\wp
_{12}^{\prime }\wp _{13}^{\prime }+288\wp _{12}^{\prime }\wp _{23}^{\prime
}+18\wp _{13}^{\prime }\wp _{23}^{\prime }%
\end{array}
\nonumber \\
&& 
\begin{array}{l}
-5760\wp _{12}^{3}-648\wp _{13}^{3}-288\wp _{23}^{3}-\wp _{12}^{2}(3834\wp
_{13}+1350\wp _{23})%
\end{array}
\nonumber \\
&& 
\begin{array}{l}
+\wp _{13}^{2}(594\wp _{12}-594\wp _{23})-\wp _{23}^{2}(648\wp _{12}-324\wp
_{13})%
\end{array}
\nonumber \\
&& 
\begin{array}{l}
+g_{2}\left( \frac{5085}{2}\wp _{12}+\frac{2061}{2}\wp _{13}+990\wp
_{23}\right)%
\end{array}
\nonumber
\end{eqnarray}
\textit{and }$I^{\ast }$\textit{\ is the operator adjoined to }$I.$

\textit{The integral }$L_{4}=L_{13}+\frac{1}{2}L_{23},$\textit{\ where }$%
L_{23}$\textit{\ is given by the same formula~(\ref{5}) after permutation of 
}$x_{1}$\textit{\ and }$x_{2}$\textit{\ and }$\partial _{1}$\textit{\ and }$%
\partial _{2},$\textit{\ is an additional integral which together with }$%
L_{1},$\textit{\ }$L_{2}, $\textit{\ }$L_{3}$\textit{\ guarantees the
algebraic integrability.}

Let us first comment on how this new integral $L_{13}$ has been found. The
highest symbol has been borrowed from the trigonometric case \cite{VFC}. The
commutativity relation between this integral and the Hamiltonian $L_{1}$
imposes a very complicated overdetermined system of relations on the
coefficients of the integral. We have resolved these relations combining the
direct analysis with the use of a computer. The addition theorem and the
differential equations for the elliptic $\wp -$function play the essential
role in these calculations. The fact that this overdetermined system has a
solution seems to be remarkable.

It is obvious from the explicit formula that $L_{13}$ commutes with $L_{2}.$
The commutativity of $L_{13}$ and $L_{3}$ has been checked with the help of
a computer. We have used a special program, which has been created for this
purpose, and the same technical tricks as in our previous paper~\cite{KhP}.
The commutativity of the operators $L_{1},$ $L_{2},$ $L_{3}$ has been proved
in~\cite{Kh}.

It is easy to check that the highest symbol of $L_{4}$ takes different
values on the solutions of the corresponding system~(\ref{odin}). This
completes the proof of the algebraic integrability.

\textit{Remark.} We should mention that according to Krichever's general
result (see \cite{Kr}) integrals $L_{1},$ $L_{2},$ $L_{3},L_{4}$ satisfy
certain algebraic relations (spectral relations). In \cite{KhP} we have
found explicitly these relations in the non-deformed case $m=1.$ In the
deformed case it seems to be a much more involved problem.

The authors are grateful to O. A. Chalykh and A. P. Veselov, who attracted
our attention to this problem.


\begin{thebibliography}{9}
\bibitem{Kr} {\scriptsize Krichever I.M., Methods of Algebraic Geometry in
the Theory of Nonlinear Equations, \textit{Uspekhi Mat. Nauk, }1977, V.32, N
6, 198--245.}

\bibitem{ChV} {\scriptsize Chalykh O.A. and Veselov A.P., Commutative Rings
of Partial Differential Operators and Lie Algebras, \textit{Commun. Math.
Phys.}, 1990, N 126, 597--611.}

\bibitem{VSCh} {\scriptsize Chalykh O.A., Styrkas K.L. and Veselov A.P.,
Algebraic Integrability for Schr\"{o}dinger Equations and Finite Reflection
Groups, \textit{Theor. Math. Phys., 1993, }V.94, N 2, 253--275.}

\bibitem{VFC} {\scriptsize Chalykh O.A., Feigin M.V. and Veselov A.P., New
Integrable Deformation of Quantum Calogero--Moser Problem, \textit{Usp. Mat.
Nauk} 1996, V.51, N 3, 185-186.}

\bibitem{CMR} {\scriptsize Calogero F., Marchioro C. and Ragnisco O., Exact
Solution of the Classical and Quantal One-Dimensional Many-Body Problems
with the Two-Body Potential }$V_{a}(x)=g^{2}a^{2}/sh^{2}ax${\scriptsize , 
\textit{Lett.Nuovo Cim.1975, }V.13, N 10, 383--387.}

\bibitem{Kh} {\scriptsize Khodarinova L.A., On Quantum Elliptic
Calogero-Moser Problem, \textit{Vestnik Mosc. Univ.}, Ser. Math. and Mech.,
1998, V.53, N 5, 16-19.}

\bibitem{BEG} {\scriptsize Braverman A., Etingof P. and Gaitsgory D.,
Quantum Integrable Systems and Differential Galois Theory, \textit{%
Transform.Groups.} 1997, V.2, N 1, 31-56.}

\bibitem{KhP} {\scriptsize Khodarinova L.A.\ and Prikhodsky I.A., Algebraic
Spectral Relations for Elliptic Quantum Calogero-Moser Problems, \textit{J.
of Nonlin. Math. Phys.}, 1999, V.6, N 3, 263-268.}
\end{thebibliography}
\end{document}